\documentclass[twocolumn,prl,floatfix,superscriptaddress]{revtex4-1}

\usepackage{amsmath}
\usepackage{amssymb}
\usepackage{graphicx}
\usepackage{color}

\newcommand{\vsd}{$V_\text{sd}${} }
\newcommand{\vsdm}{V_\text{sd}}

\begin{document}

\title{Quantized conductance doubling and hard gap in a two-dimensional semiconductor-superconductor heterostructure}

\author{M. Kjaergaard}
\affiliation{Center for Quantum Devices and Station Q Copenhagen, Niels Bohr Institute, University of Copenhagen, Universitetsparken 5, 2100 Copenhagen, Denmark}
\author{F.~Nichele}
\affiliation{Center for Quantum Devices and Station Q Copenhagen, Niels Bohr Institute, University of Copenhagen, Universitetsparken 5, 2100 Copenhagen, Denmark}
\author{H.~J.~Suominen}
\affiliation{Center for Quantum Devices and Station Q Copenhagen, Niels Bohr Institute, University of Copenhagen, Universitetsparken 5, 2100 Copenhagen, Denmark}
\author{M.~P.~Nowak}
\affiliation{Kavli Institute of Nanoscience, Delft University of Technology, P.O. Box 4056, 2600 GA Delft, The Netherlands}
\affiliation{QuTech, Delft University of Technology, P.O. Box 4056, 2600 GA Delft, The Netherlands}
\affiliation{AGH University of Science and Technology, Faculty of Physics and Applied Computer Science, al. Mickiewicza 30, 30-059 Krak\'{o}w, Poland}
\author{M.~Wimmer}
\affiliation{Kavli Institute of Nanoscience, Delft University of Technology, P.O. Box 4056, 2600 GA Delft, The Netherlands}
\affiliation{QuTech, Delft University of Technology, P.O. Box 4056, 2600 GA Delft, The Netherlands}
\author{A.~R.~Akhmerov}
\affiliation{Kavli Institute of Nanoscience, Delft University of Technology, P.O. Box 4056, 2600 GA Delft, The Netherlands}
\author{J.~A.~Folk}
\affiliation{Department of Physics and Astronomy, University of British Columbia, Vancouver, British Columbia, V6T1Z1, Canada}
\affiliation{Quantum Matter Institute, University of British Columbia, Vancouver, BC, V6T1Z4, Canada}
\author{K.~Flensberg}
\affiliation{Center for Quantum Devices and Station Q Copenhagen, Niels Bohr Institute, University of Copenhagen, Universitetsparken 5, 2100 Copenhagen, Denmark}
\author{J.~Shabani}\thanks{Now at City College, City University of New York}
\affiliation{California NanoSystems Institute, University of California, Santa Barbara, CA 93106, USA}
\author{C.~J.~Palmstr\o{}m}
\affiliation{California NanoSystems Institute, University of California, Santa Barbara, CA 93106, USA}
\author{C.~M.~Marcus}\thanks{Corresponding author. Electronic address: marcus@nbi.dk}
\affiliation{Center for Quantum Devices and Station Q Copenhagen, Niels Bohr Institute, University of Copenhagen, Universitetsparken 5, 2100 Copenhagen, Denmark}

\date{\today}

\maketitle
%
%
\textbf{Coupling a two-dimensional (2D) semiconductor heterostructure to a superconductor opens new research and technology opportunities, including fundamental problems in mesoscopic superconductivity, scalable superconducting electronics, and new topological states of matter. One route towards topological matter is by coupling a 2D electron gas with strong spin-orbit interaction to an s-wave superconductor. Previous efforts along these lines have been adversely affected by interface disorder and unstable gating. Here we show measurements on a gateable InGaAs/InAs 2DEG with patterned epitaxial Al, yielding devices with atomically pristine interfaces between semiconductor and superconductor. Using surface gates to form a quantum point contact (QPC), we find a hard superconducting gap in the tunneling regime. When the QPC is in the open regime, we observe a first conductance plateau at $4e^2/h$, consistent with theory. The hard-gap semiconductor-superconductor system demonstrated here is amenable to top-down processing and provides a new avenue towards low-dissipation electronics and topological quantum systems.}

%
%
\textbf{Introduction}---Recent work on semiconductor nanowires has offered evidence for the existence of Majorana zero modes, a signature of the topological superconductivity \cite{Mourik:2012je,Das:2012hi,Deng:2012gn}. A characteristic of the first studies in this area was significant subgap tunneling conductance (a so-called soft gap), attributed to disorder at the semiconductor-superconductor (Sm-S) interface \cite{Takei:2013id,Cole:2015bl}. In nanowires, the soft-gap problem was recently resolved by growing Al epitaxially on InAs nanowires, yielding greatly reduced subgap conductance \cite{Chang:2015kw,Higginbotham:2015ii}. Studies of Sm-S systems based on top-down processed gateable two-dimensional electron gases (2DEGs) coupled to superconductors have not explicitly addressed the soft-gap issue yet \cite{Amado2013,Irie2014}. However experiments on such systems have demonstrated other theoretical predictions, such as quantization of critical current \cite{Irie2014,Takayanagi:1995hg,Bauch:2005gh}, the retro-reflection property of Andreev scattering \cite{Jakob:2000cw}, and spectroscopy of a gate-defined quantum dot with superconducting leads \cite{Deon2011,Deon2011_2}, which do not require a hard proximity-induced gap in the semiconductor. 
\begin{figure}[!ht]
\center
\includegraphics[width = 8.8 cm]{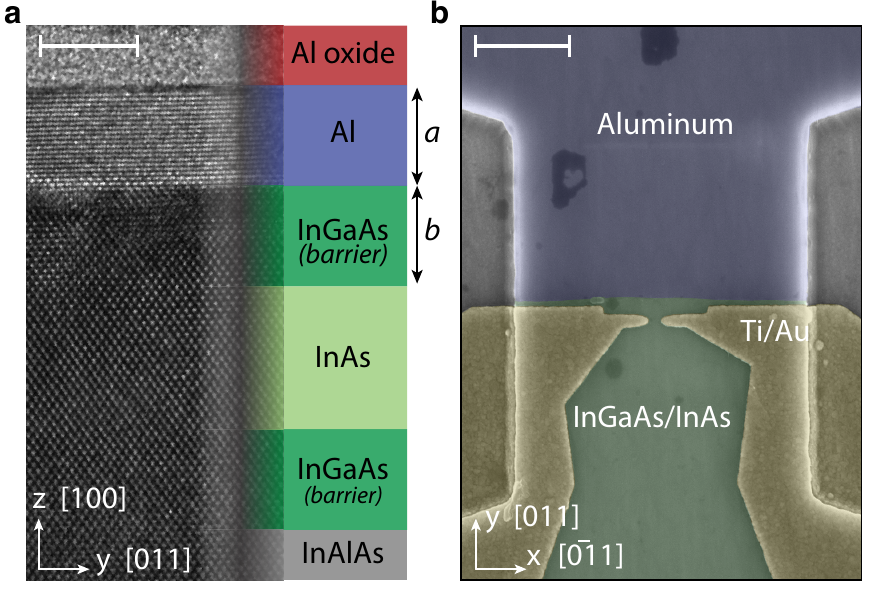}
\caption{\footnotesize{\textbf{Epitaxial aluminium on InGaAs/InAs and device layout}. 
\textbf{a}, Cross-sectional transmission electron micrograph of epitaxial Al on InGaAs/InAs. On the wafer imaged here, the height of the InGaAs barrier is $b=5~$nm and Al film thickness $a \sim 5~$nm. The scalebar is $5~\text{nm}$. \textbf{b}, False-color scanning electron micrograph of Device 1 (see main text for details). The scalebar is $1~\mu\text{m}$.}}
\label{fig1}
\end{figure}

The two main results we present in this paper are both consequences of the transparent epitaxial Sm-S interface and overcome the soft gap problem for 2D electron gases. The first is a doubling of the the lowest quantized conductance plateau, from $2e^{2}/h$ in the normal state to $4e^{2}/h$ in the superconducting state, as predicted theoretically \cite{Beenakker:1992dd}. The second is a strong suppression of conductance for voltages smaller than the superconducting gap when the QPC is in the tunneling regime---that is, the detection of a hard superconducting gap in a proximitized 2DEG. Conductance doubling arises from Andreev reflection transferring charge $2e$ into the superconductor \cite{Andreev:1964uy}. The hard gap reflects the absence of electronic states below the superconducting gap in the semiconductor. Using gate voltage to control the QPC, we measure conductance across the transition from weak tunneling to the open-channel regime and find good (but not perfect) agreement with the theory of a normal-QPC-superconductor structure \cite{Beenakker:1992dd}.

\textbf{Results}\\
\emph{Properties of the 2DEG and the superconducting Al film}---The starting material is an undoped InAs/InGaAs heterostructure with epitaxial Al as a top layer, grown by molecular beam epitaxy \cite{Javad:Xwv0duEK}. A cross-sectional TEM showing a sharp epitaxial Sm-S interface is shown in Fig.~\ref{fig1}a. In the devices reported here, the thickness of the InGaAs barrier was $b=10~$nm, and the Al film was $a=10~$nm. A Hall ball fabricated on the same wafer with the Al removed (see Methods) gave density $n = 3\cdot 10^{12}~$cm$^{-2}$ and mobility $\mu = 10^4~$cm$^2\text{V}^{-1}\text{s}^{-1}$, yielding a mean free path $l_\textrm{e} \sim 230~\text{nm}$. In a similar wafer, weak anti-localization analysis gave a spin-orbit length $l_\text{so} \sim 45~\text{nm}$  \cite{Javad:Xwv0duEK}. The Al film has a critical temperature $T_\mathrm{c} = 1.56~$K, corresponding to a gap $\Delta_0 = 235~\mu\text{eV}$, enhanced from the bulk value of Al, and consistent with other measurements on Al films of similar thickness \cite{Chubov:1969uu}. The in-plane critical field of the Al film is $B_\mathrm{c} = 1.65~$T \cite{Javad:Xwv0duEK}.

\begin{figure}[!ht]
\center
\includegraphics[width = 8.8 cm]{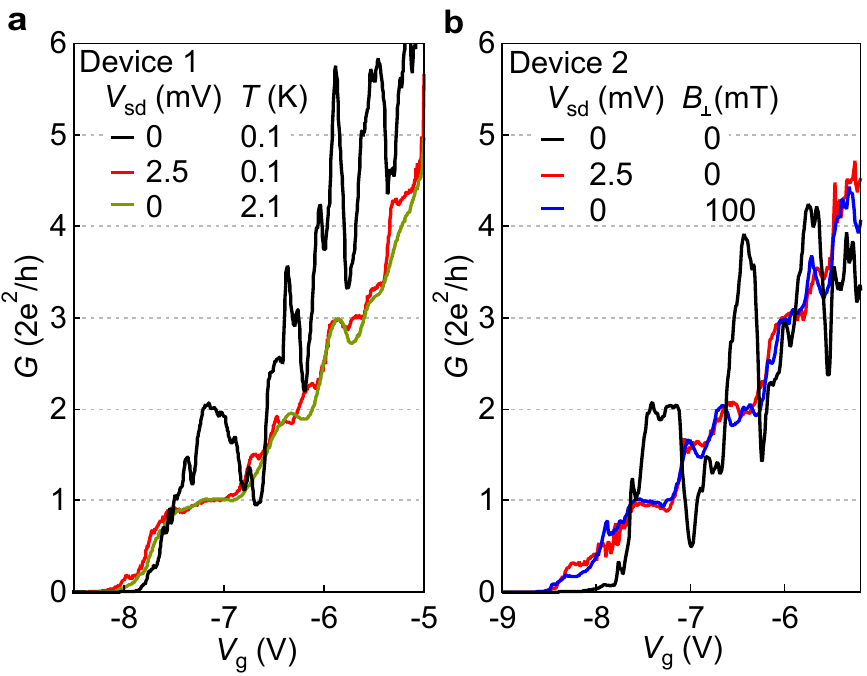}
\caption{\footnotesize{\textbf{Quantized conductance in the Andreev quantum point contact}}. \textbf{a}, Differential conductance, $G$, as a function of gate voltage $V_\textrm{g}$ at zero bias (black line), at source-drain bias larger than the gap (red line), and at elevated temperature (green line). At zero bias and base temperature, the first conductance plateau is at $4e^2/h$, double the value at higher temperature or bias. \textbf{b}, The differential conductance in a second, lithographically identical, device at zero bias (black line), at source-drain bias larger than the gap (red line), and in a magnetic field applied perpendicular to the plane of the chip (blue line).}
\label{fig2}
\end{figure}

\emph{Quantized conductance doubling}---A scanning electron micrograph of Device 1 is shown in Fig.~\ref{fig1}b.The conductance of the QPC is tuned by negative voltages applied to the gates. The QPC is located $\sim 150~\text{nm}$ in front of the region where the Al film has not been removed.  Figure \ref{fig2} shows conductance traces for the two lithographically similar QPCs. In the superconducting state, both devices show increased conductance at the plateau of the QPC and suppressed conductance below $G \sim 0.8G_0$, where $G_{0}\equiv 2e^{2}/h$, relative to the normal state. This behavior is the hallmark of Andreev reflection being the dominant conduction mechanism through the QPC \cite{Beenakker:1992dd,Mortensen:1999gw}. Raising the temperature above the critical temperature of the Al film, applying an out-of-plane magnetic field, or applying a bias larger than the gap, all bring the lowest plateau back to $2e^2/h$  (see Fig.~\ref{fig2}). The dip structure at the transition between conductance plateaus was also observed in a similar experiment on nanowires \cite{Zhang:2016vl}, and is presumably caused by mode mixing due to disorder, leading to a reduction in transparency of the already open first channel. A constant contact resistance $R_\textrm{c} \sim 1~$k$\Omega$ has been subtracted in each viewgraph, a value chosen to move the first plateau in the normal state to $G_0$.

\begin{figure*}[!ht]
\centering
\includegraphics[scale=0.8]{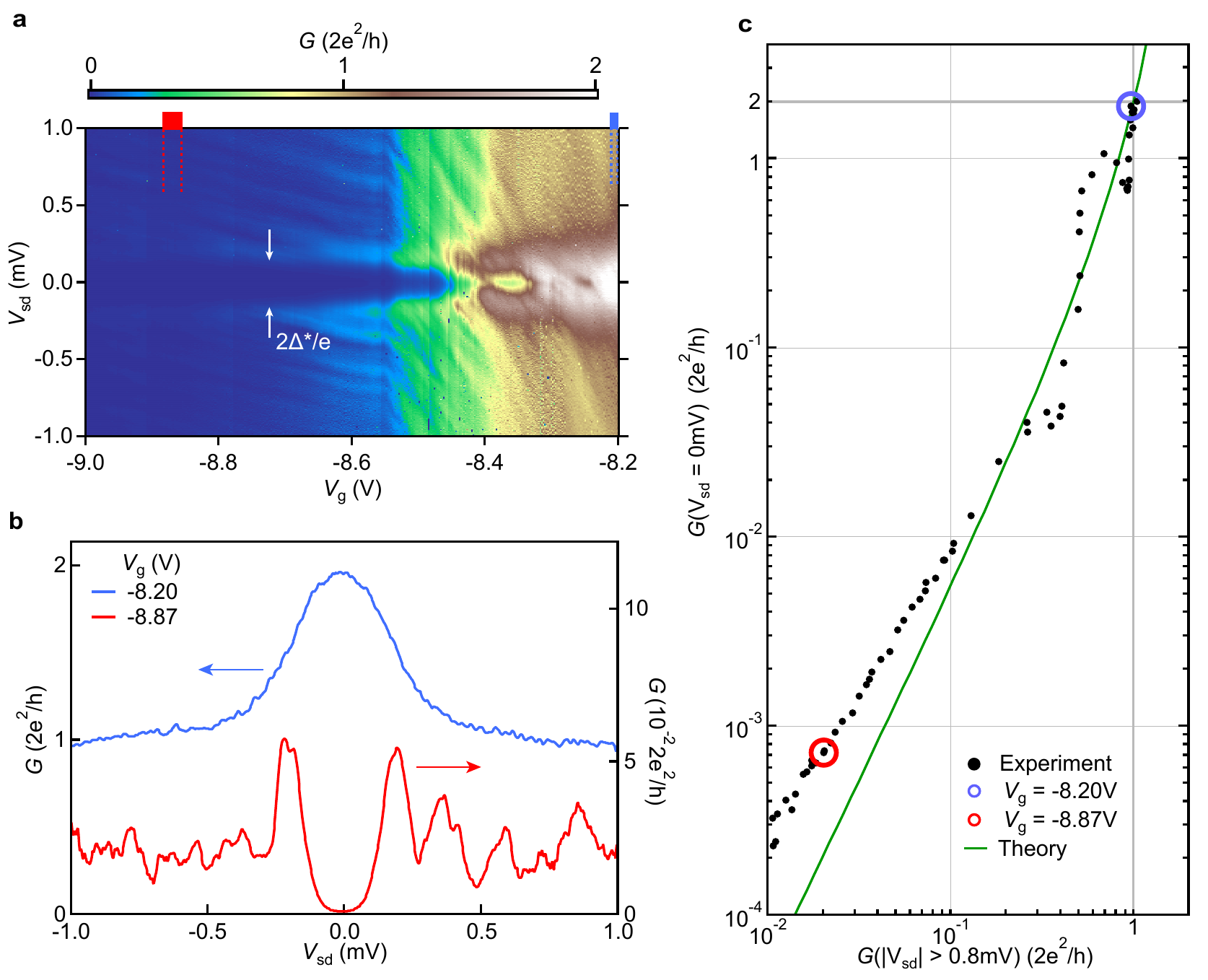}
\caption{\footnotesize{\textbf{Transition from $4e^2/h$ conductance to hard superconducting gap.} \textbf{a}, Differential conductance, $G$, in Device 1 as a function of gate voltage $V_\text{g}$ and source-drain voltage bias $V_\text{sd}$. \textbf{b}, Vertical cuts in \textbf{a} in the tunneling (red line) and one-channel (blue line) regime. Supplementary Fig. 1 shows data from a lithographically similar device on a wafer with no InGaAs barrier (i.e. $b=0~\text{nm}$) between the top layer Al and the InAs 2DEG. \textbf{c}, Differential conductance at zero source-drain voltage, $G(V_\text{sd} = 0~\text{mV})$, versus averaged differential conductance at finite source-drain voltage, $G(|V_\text{sd}| > 0.8~\text{mV})$. Red and blue circles indicate data corresponding to cuts in \textbf{b}. Green line is theoretical prediction for conductance in an Andreev enhanced QPC (Eq. \eqref{eq:SNconductance} with no fitting parameters).}}
\label{fig3}
\end{figure*}

\emph{Hard superconducting gap}---By further depleting the electron gas in the constriction, the device is operated as a tunnel probe of the local density of states in the InAs 2DEG. This technique has been applied to studying subgap properties of semiconductor nanowires coupled to superconductors \cite{Mourik:2012je,Deng:2012gn,Das:2012hi,Churchill:2013cq,Lee:2014gj,Chang:2015kw}. In Fig.~\ref{fig3}a the QPC voltage is decreased to gradually transition from the one-channel regime, where the zero bias conductance is $4e^2/h$, to the tunneling regime, where conductance is strongly suppressed for $|V_\text{sd}| < 190~\mu\text{V}$. From these measurements, the gap in the density of states of the InAs due to the proximity to the Al is estimated to be $\Delta^\star \sim 190~\mu\text{eV}$ (measured peak-to-peak). The value of $\Delta^\star$ is similar, but not identical, to the gap in the Al film as estimated from $T_\textrm{c}$, as discussed above.

In the case of perfect Andreev reflection from the superconductor/semiconductor interface, the conductance of one channel through a constriction proximal to the interface is given by 
\begin{equation}
G_\text{ns} = 2G_0\frac{G_\text{nn}^2}{(2G_0 - G_\text{nn})^2} \label{eq:SNconductance},
\end{equation}
where $G_\text{ns}$ is the conductance when the film is superconducting, and $G_\text{nn}$ is the conductance in the normal state \cite{Beenakker:1992dd}. In Fig.~\ref{fig3}c the prediction in Eq.~\eqref{eq:SNconductance} with no free parameters (green line) and experimental data are shown. Here, $G_\text{nn}$ is the average conductance for $|\vsdm| > 0.8~\text{mV}$, justified by the equality of applying a bias and raising the temperature above $T_\text{c}$, as shown in Fig.~\ref{fig2}a.  Equation \eqref{eq:SNconductance} is consistent with the data over two orders of magnitude in $G_\text{ns}$, indicating that the zero bias conductance up to $4e^2/h$ is well described by the prediction of perfect Andreev reflection of a single QPC mode. Equation \eqref{eq:SNconductance} represents the only quantitative theory of the relation between subgap conductance and normal state conductance (i.e. the hard gap) of which we are aware, and the agreement between Eq.~\eqref{eq:SNconductance} and the experiment in Fig.~\ref{fig3}c leads to the designation of a hard gap in this superconductor-2DEG system. However, the systematic deviation between data and prediction in Fig.~\ref{fig3}c for $G_\text{ns} < 10^{-2}~2e^2/h$ could be a manifestation of a small remnant non--zero normal scattering probability.

The shapes of the conductance curves at $e\vsdm \lesssim \Delta^\star$ in the tunneling regime (red line in Fig.~\ref{fig3}b) are smeared relative to the conventional Bardeen-Cooper-Schrieffer (BCS) density of states of a superconductor. This could be due to broadening of the BCS coherence peaks in the disordered superconducting film formed in the 2DEG under the Al \cite{Feigelman:2012fp}, a weak coupling between Al and 2DEG \cite{Cole:2015bl} or the layout of the tunnel probe relative to the proximitized 2DEG \cite{Gueron:1996hw,leSueur:2008kl,Cherkez:2014jz}.

\begin{figure*}[ht]
\centering
\includegraphics[scale=0.8]{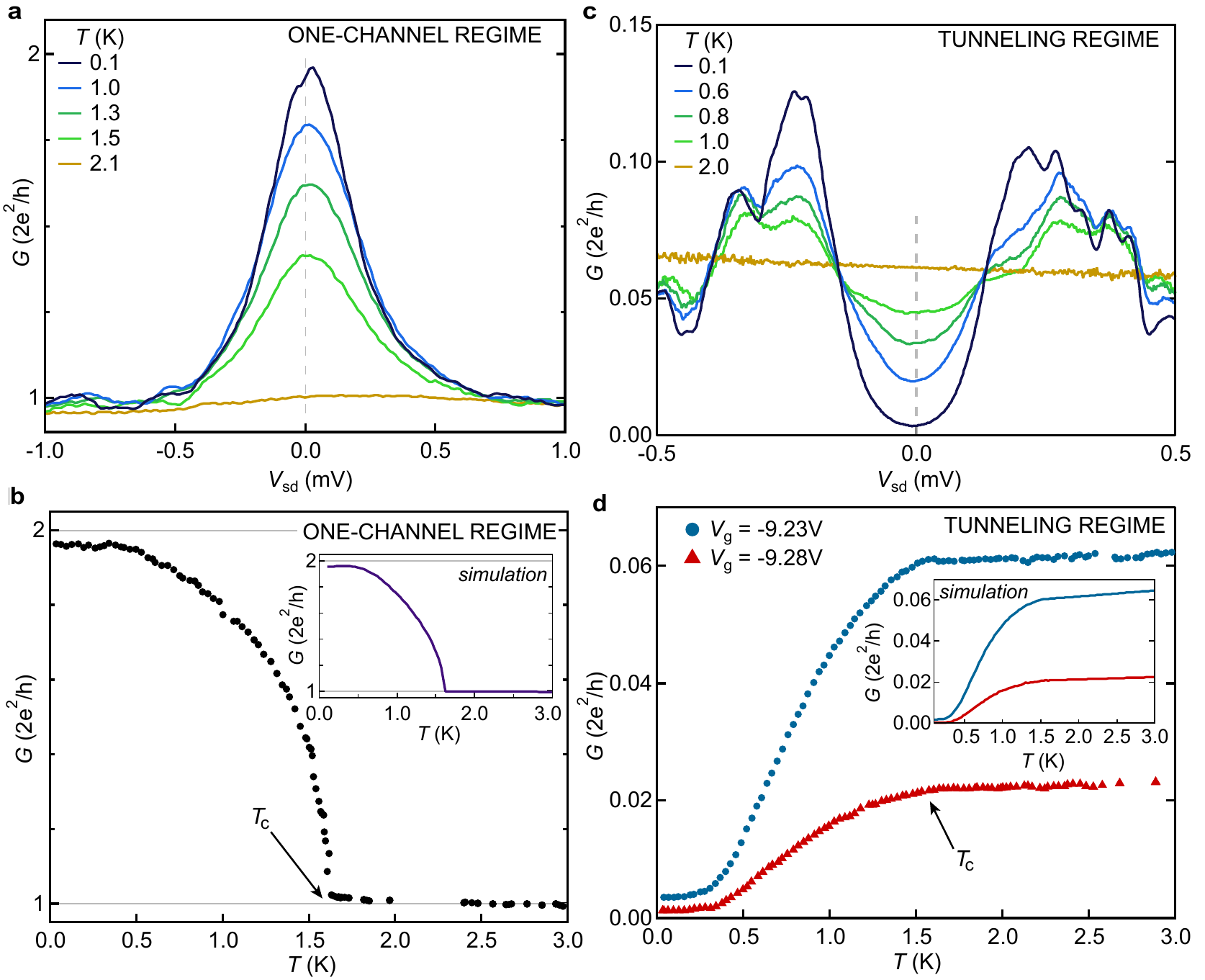}
\caption{\footnotesize{\textbf{Temperature dependence of the enhanced subgap conductance and the hard superconducting gap.} 
\textbf{a}, Differential conductance, $G$, as a function of source-drain bias voltage, $V_\text{sd}$, at five temperatures in the one-channel regime. See Supplementary Fig. 2a for similar data measured on a wafer with no InGaAs barrier between the top layer Al and the InAs 2DEG.
\textbf{b}, Temperature dependence at zero bias (corresponding to cut along the dashed, gray line in \textbf{a}) in the one-channel regime.
\textbf{c}, Similar measurement to \textbf{a}, but in the tunneling regime.
\textbf{d}, As in \textbf{b}, for two different values of gate voltage, $V_\textrm{g}$, both in the tunneling regime. Insets in \textbf{b} and \textbf{d} show results from numerical simulations (see Supplementary Figures 3, 4 and 5 for more details on numerical results).
}}
\label{fig4}
\end{figure*}

\emph{Temperature dependence of the density of states}---The temperature dependence of the conductance in the Andreev QPC is different in the one-channel and in the tunnel regime (Fig.~\ref{fig4}). The one-channel regime (Fig.~\ref{fig4}a,b) has a pronounced kink at $T = T_\mathrm{c}$, presumably associated with the sudden onset of Andreev enhanced subgap conductance. In contrast, the temperature-dependence in the tunnel regime (Fig.~\ref{fig4}c,d) is smeared close to $T_\textrm{c}$ due to thermally excited quasiparticles. 

The temperature dependence is simulated (insets in Fig.~\ref{fig4}) by calculating $G = \int dE \mathcal G(E) (-\frac{\partial f}{ \partial E})$ where $f$ is the Fermi function which accounts for thermal broadening. The conductance $\mathcal G(E)$ is calculated by combining scattering matrices of quasielectrons and quasiholes in the normal region and Andreev reflection at the superconductor interface (details of the simulation is given in methods). The scattering matrices are calculated using the numerical package Kwant \cite{Groth:2014ia}, and the simulation is performed using the device geometry from the micrograph in Fig.~\ref{fig1}b. The temperature dependence of the gap is modeled with $\Delta^\star(T) = \Delta^\star \sqrt{1-(T/T_\textrm{c})^2}$, and the Andreev reflection amplitude is taken from \cite{Beenakker:1992dd}. The simulation shows good quantitative agreement with the data. 

\begin{figure}[!ht]
\center
\includegraphics[width = 8.8cm]{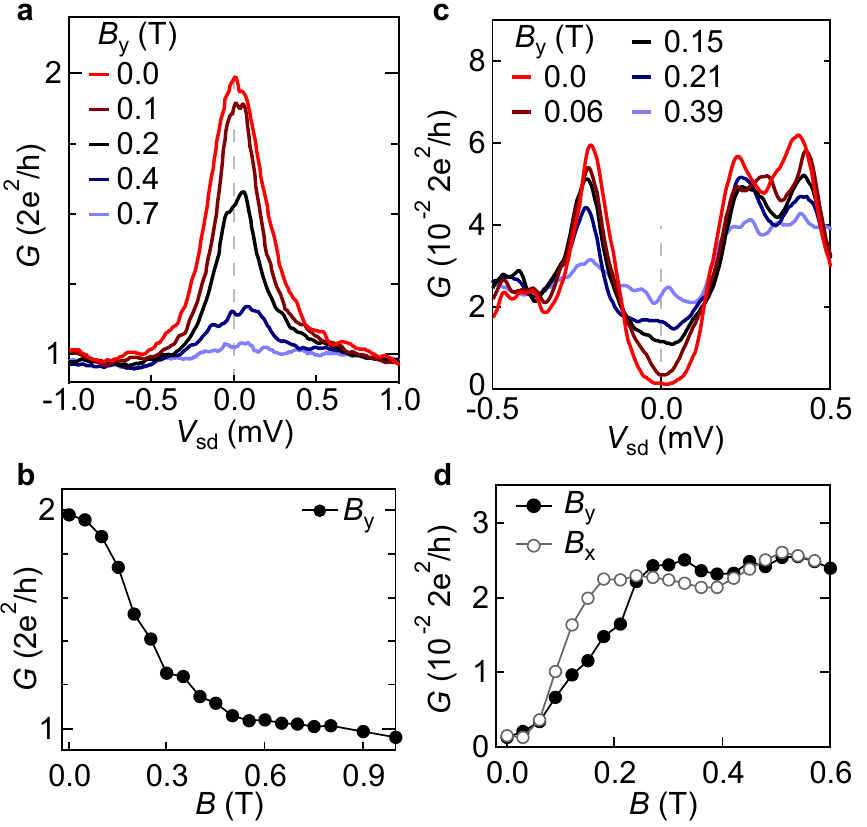}
\caption{\footnotesize{\textbf{In plane magnetic field of the enhanced subgap conductance and the hard superconducting gap.} 
\textbf{a}, Differential conductance, $G$, as a function of source-drain bias, $V_\text{sd}$, at several in-plane magnetic fields applied along the point contact constriction.
\textbf{b}, Zero-bias conductance as a function of the in-plane magnetic field, $B_y$.
\textbf{c}, Similar measurement to \textbf{a} but in the tunneling regime. Supplementary Fig. 2b show data on a lithographically similar device fabricated on a wafer with no InGaAs barrier between the top layer Al and InAs 2DEG (i.e. $b = 0~\text{nm}$).
\textbf{d}, As in \textbf{b}, but in the tunneling regime, for both in-plane directions ($B_y$ is along and $B_x$ is perpendicular to the constriction).}}
\label{fig5}
\end{figure}

\emph{Magnetic field dependence of the density of states}---To drive a superconductor/semiconductor device into a topological regime, one requirement is $g\mu_\textrm{B} B > \Delta^\star$, while the native superconductor retains its gap. Figure~\ref{fig5} shows the in-plane magnetic field dependence of $\Delta^\star$, from which an approximate critical field $B_\textrm{c}^\star \sim 300~$mT is extracted. A rough estimate of the $g$-factor can be inferred by assuming the critical $B_\textrm{c}^\star$ results from Zeeman energy surpassing the induced superconducting gap, that is $g\mu_\textrm{B} B_\textrm{c}^\star= \Delta^\star$, which yields $g \sim 10$, similar to the $g$-factor in bulk InAs. In Fig.~\ref{fig5}d the zero-bias conductance is shown for the two different in-plane directions, and the slight direction dependence of $B_\textrm{c}^\star$ could be due to an anisotropic $g$-factor in the InAs crystal lattice. The induced gap in the 2DEG disappears at in plane magnetic fields significantly smaller than the critical field of the Al film itself. The 2DEG has a strong spin-orbit interaction ($l_\text{so} \sim 45~\text{nm}$), which, taken together with the intimate coupling to the superconductor, makes this material system a feasible candidate to realize topological superconducting devices. By using top-down fabrication techniques and the electrostatic gating demonstrated here, effective 1D systems can be produced, in which an in-plane magnetic field can close the induced superconducting gap to reach a topological phase.

In conclusion, we observe quantization doubling through a QPC proximal to a superconductor/semiconductor interface, confirming a long-standing theoretical prediction \cite{Beenakker:1992dd}. Operated as a gate-tunable tunnel probe of the local density of states, the QPC shows a hard superconducting gap induced in the 2DEG. The magnetic field dependence of the induced gap compares favorably with the critical field of the superconducting film, opening possibilities to pursue topological states of matter in one-dimensional structures fabricated from epitaxial Al/2D InAs material.

%
%
\vspace{0.1in}
\textbf{Methods}\\{\small 
%
%
\emph{Fabrication and measurement setup}---Ohmic contacts to the InAs electron gas are formed by the epitaxial Al directly and mesa structures are patterned by standard III-V chemical etching techniques. The aluminum is etched using commercial Transene Aluminum Etch D. Subsequent to the selective Al etch, an insulating 40 nm Al$_2$O$_3$ layer is deposited using atomic layer deposition and metallic gates (5 nm Ti/50 nm Au) are evaporated onto the device. The measurements were performed in a dilution refrigerator with a base mixing chamber temperature $T_\textrm{mc} \sim 30~$mK, using four-terminal lock-in techniques and DC measurements.

%
%
\vspace{0.1in}
\emph{Measurement details}---The data in Fig.~\ref{fig3} is measured in a DC setup, incrementing the voltage in steps of size $3 ~\mu$V. The data is smoothed over 10 steps and the derivative is calculated numerically to obtain the differential conductance. A constant contact resistance $R_\text{c} = 800~\Omega$ is subtracted from the data, moving the conductance at $V_\text{g} = -8.2~$V for $\vsdm > 0.8~$mV to $2e^2/h$. The 4-terminal resistance of the device is $R_\text{d}$ = 400$~\Omega$ with $V_\text{g} = 0~$V. The difference between $R_\text{c}$ and $R_\text{d}$ is most likely dominated by the change of resistivity near the gated region, when the gate is turned on, as well as the distance from the voltage probe to the QPC region. The voltage probes are located $\sim 15~\mu$m away from the QPC and the gates overlap the mesa over an area $\sim 1.6~\mu$m$^2$. The normal state conductance is calculated as the average of $G(\vsdm)$ for \vsd in the range $[\pm 0.8~\text{mV}, \pm 1~\text{mV}]$. The analysis is largely unaffected by changing the averaging window for values $|\vsdm| > 0.6~$mV. The cuts in Fig.~3b are taken by averaging over a 12~mV (30~mV) window in $V_\text{g}$ for the one-channel (tunneling) regime. Finally, each datapoint in Fig.~3c is calculated as the average over a $10~$mV range in $V_\text{g}$.

%
%
\vspace{0.1in}
\emph{Model for numerical simulations}---We calculate the conductance of the junction in two steps. Firstly, we determine the scattering properties of the normal region which we assume is a 1.1 $\mu$m wide channel of length $L$, where we have taken dimensions from SEM in Fig.~1b. It is described by the spinless Hamiltonian,
\begin{equation}
H = \frac{\hbar^2\mathbf{k}^2}{2m^*} + V_\text{QPC}(x, y) + V_\text{d}(x,y) - \mu.
\label{hamiltonian}
\end{equation}

We model the QPC as two rectangular gates located at $x = 400$ nm, with the width $2W$, separated by the length $2S$ and located at the distance $d$ above 2DEG (see Supplementary Fig.~3 for illustration of $W$ and $S$). We calculate the potential generated by the QPC electrodes, $V_\text{QPC}(x,y)$, for the gate voltage $V_g$ following \cite{davies_modeling_1995}, with
\begin{equation}
\begin{split}
\frac{V_\text{QPC}(x,y)}{-eV_{g}} = \frac{1}{\pi}\left[ \arctan\left(\frac{W+x}{d}\right)+\arctan\left(\frac{W-x}{d}\right) \right]\\
-g(S+y,W+x)-g(S+y,W-x)\\
-g(S-y,W+x)-g(S-y,W-x),
\end{split}
\end{equation}
where
\begin{equation}
g(u,v) = \frac{1}{2\pi}\arctan\left(\frac{uv}{dR} \right),
\end{equation}
and $R = \sqrt{u^2 + v^2 + d^2}$. The potential landscape of the simulation is shown in Supplementary Fig.~3.

We include disorder \cite{ando_quantum_1991} by adding a random on-site energy $V_\text{d}(x,y)$ distributed uniformly between $-W/2$ and $W/2$ where
\begin{equation}
W = \mu\sqrt{\frac{6\lambda_F^3}{\pi^3\Delta x^2l_e}}.
\end{equation}
Due to limitation of the computational mesh resolution we exclude the disorder from the vicinity of the QPC and take $W\neq0$ only for $x > 700$ nm.

We calculate the scattering matrix of the normal part of the junction for a quasiparticle at the energy $\varepsilon$ as
\begin{equation}
S_N(\varepsilon) =
\begin{pmatrix}
r(\varepsilon) & t(\varepsilon) \\
t'(\varepsilon) & r'(\varepsilon)
\end{pmatrix},
\end{equation}
using Kwant package \cite{Groth:2014ia} and discretizing the Hamiltonian in Eq. (\ref{hamiltonian}) on a mesh with the spacing $\Delta x = \Delta y = 3$ nm. The quantities $r(E)$ and $t(E)$ denote reflection and transmission submatrices for a time-reversal symmetric system. In the second step, we combine the scattering matrices calculated for $\varepsilon$ and $-\varepsilon$ (that correspond to quasielectron and quasihole respectively) with the matrix that accounts for the Andreev reflection at the superconductor interface
\begin{equation}
S_A = r_A
\begin{pmatrix}
0 & e^{i \phi} \\
e^{-i \phi} & 0
\end{pmatrix},
\end{equation}
where
\begin{equation}
\quad r_A = \frac{\varepsilon}{\Delta(T)} -i\ \textrm{sign}\left[\varepsilon+\Delta(T)\right]\sqrt{1 - \frac{\varepsilon^2}{\Delta(T)^2}}.
\end{equation}
The latter equation describes the Andreev reflection amplitude \cite{Beenakker:1992dd} including the temperature dependent pairing potential $\Delta(T) = \Delta^* \sqrt{1-(T/T_\text{c})^2}$. Finally we calculate the conductance according to
\begin{equation}
G_\text{ns}(E) = \int d\varepsilon \mathcal G(\varepsilon) \left(-\frac{\partial f(E,\varepsilon)}{ \partial \varepsilon}\right),
\end{equation}
where $f$ stands for the Fermi function 
\begin{equation}
f(E, \varepsilon) = \frac{1}{e^{(\varepsilon - E)/k_\text{B}T}+1},
\end{equation}
and where $\mathcal G(\varepsilon) = N - \lVert r_e(\varepsilon) \rVert^2 + \lVert r_h(\varepsilon) \rVert^2$.
$N$ is the number of modes in the normal lead. The quasielectron and quasihole reflection matrices are given by:
\begin{equation}
r_e(\varepsilon) = r(\varepsilon) + t'(\varepsilon)\; r_A r'^*(-\varepsilon)\; r_A \frac{1}{1 - r'(\varepsilon)\; r_A r'^*(-\varepsilon)\; r_A} t(\varepsilon),
\end{equation}
\begin{equation}
r_h(\varepsilon) = t'^*(-\varepsilon)\; r_A \frac{1}{1 - r'(\varepsilon)\; r_A r'^*(-\varepsilon)\; r_A} t(\varepsilon).
\end{equation}
Additionally, the normal-state conductance is given by $G_{nn} = \lVert t(\varepsilon=0) \rVert^2$. Results of the simulations are shown Supplementary Figures~3,~4 and~5.

\vspace{0.1in}
\emph{Data and code availability statement}---All data presented in the main paper and supplement, as well as code used to generate simulations are available from the authors upon request.

}

\vspace{0.1in}
\emph{Acknowledgements:} Research support by Microsoft Project Q, the Danish National Research Foundation. C.M.M. acknowledges support from the Villum Foundation. F.N. acknowledges support from a Marie Curie Fellowship (No. 659653). M.P.N. acknowledges support from ERC Synergy Grant. A.A. is supported by an ERC Starting Grant. M.W. and A.A. are supported by the Foundation for Fundamental Research on Matter (FOM) and the Netherlands Organization for Scientific Research (NWO/OCW) as part of the Frontiers of Nanoscience program. We are indebted to S. Kraemer for the TEM analysis, performed at the UCSB MRL Shared Experimental Facilities (NSF DMR 1121053), a member of the NSF-funded Materials Research Facilities Network.

\vspace{0.1in}
\emph{Authors Contribution:} M.K., F.N., H.J.S. and C.M. conceived the experiment. M.K., F.N., H.J.S. designed, fabricated and measured the devices and wrote the manuscript, with comments from all other authors. J.A.F. and K.F. provided input on interpretations. M.P.N, M.W and A.R.A. developed theory and code for the simulations. The wafer was grown by J.S. and C.J.P.

\vspace{0.1in}
\emph{Competing interests:} The authors declare no competing interests.

%
%

\clearpage
\onecolumngrid
\begin{center}
\textbf{\large Supplementary material for ``Quantized conductance doubling and hard gap in a two-dimensional semiconductor-superconductor heterostructure''}
\end{center}
\appendix

\setcounter{equation}{0}
\setcounter{figure}{0}
\setcounter{table}{0}
\makeatletter
\renewcommand{\theequation}{S\arabic{equation}}
\renewcommand{\thefigure}{S\arabic{figure}}
\renewcommand{\bibnumfmt}[1]{[S#1]}
\renewcommand{\citenumfont}[1]{S#1}

\section{Supplementary note 1: Measurements on alternate wafer}
\begin{figure}[!ht]
\center
\includegraphics[width = 15 cm]{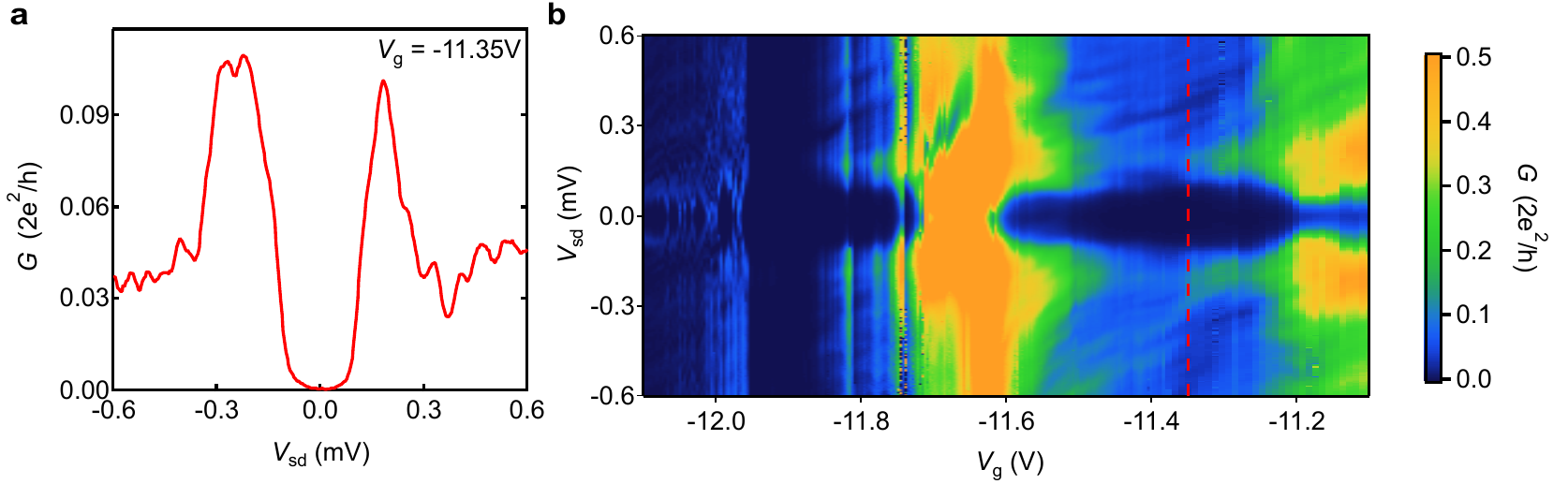}
\caption{\footnotesize{\textbf{Spectroscopy of the superconducting gap in a wafer with 0~nm InGaAs barrier}. 
\textbf{a}, Differential conductance, $G$, as a function of source-drain voltage, $V_\text{sd}$, in a quantum point contact geometry, with gate voltage $V_\text{g} = -11.35~$V. \textbf{b}, Differential conductance at finite source-drain voltage, as the split-gate is used to deplete the 2DEG by decreasing $V_\text{g}$. Vertical cut in \textbf{a} indicated by dashed, red line.}}
\label{figS1}
\end{figure}

Under identical growth conditions, a wafer without an InGaAs top barrier (i.e. $b = 0~$nm) between the epitaxial aluminum and the InAs quantum well was produced. The density and mobility, measured using a conventional Hall bar geometry, was $n = 4.5\cdot 10^{16}~$m$^{-2}$ and $\mu = 4000~$cm$^2/$Vs, corresponding to a mean free path of $l_e = 150~$nm. In a lithographically similar device to that shown in Fig.~1 of the main text, we observe a hard superconducting gap (Supplementary Figure ~\ref{figS1}). When the gates are operated in the quantum point contact regime, we did not observe quantized steps in conductance. The non-monotonic decrease in conductance at $\vsdm =0~$mV, believed to be due to disorder in the 2DEG, makes the identification of a superconducting gap in this wafer difficult (Supplementary Fig.~\ref{figS1}b).

\begin{figure}[!ht]
\center
\includegraphics[width = 15 cm]{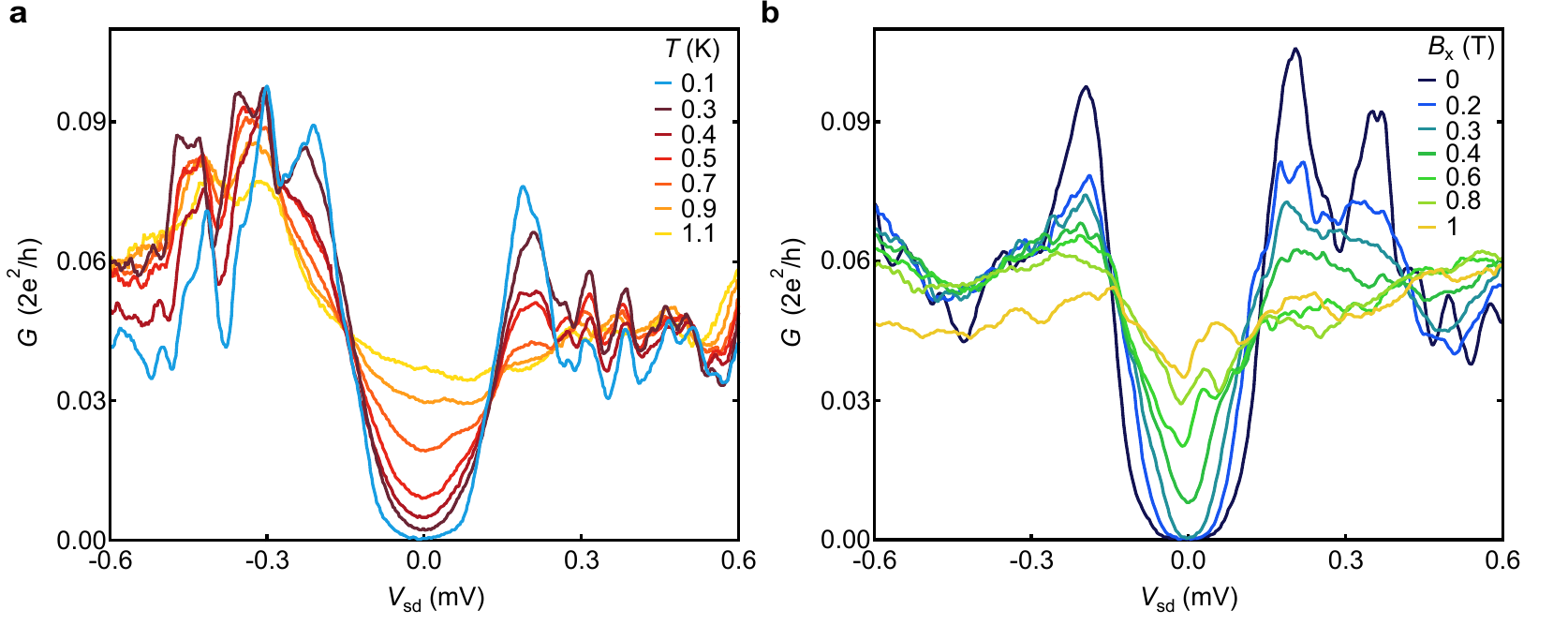}
\caption{\footnotesize{\textbf{Temperature and magnetic field dependence of the proximity induced superconducting gap}. 
\textbf{a}, Differential conductance as a function of source-drain voltage for several temperatures. \textbf{b}, In-plane magnetic field dependence of the superconducting gap (field applied perpendicular to the constriction).}}
\label{figS2}
\end{figure}
However, by increasing the temperature (Supplementary Fig.~\ref{figS2}a) or the magnetic field (Supplementary Fig.~\ref{figS2}b) we confirm that the gap in the density of states in Supplementary Fig.~\ref{figS1} is related to the superconducting properties, and not a spurious quantum dot.

\subsection{Supplementary Note 2: Numerical results}
The potential landscape generated by the simulation, used to model the QPC is shown in Supplementary Figure~3.
\begin{figure}[h!]
\center
\includegraphics[width = 13cm]{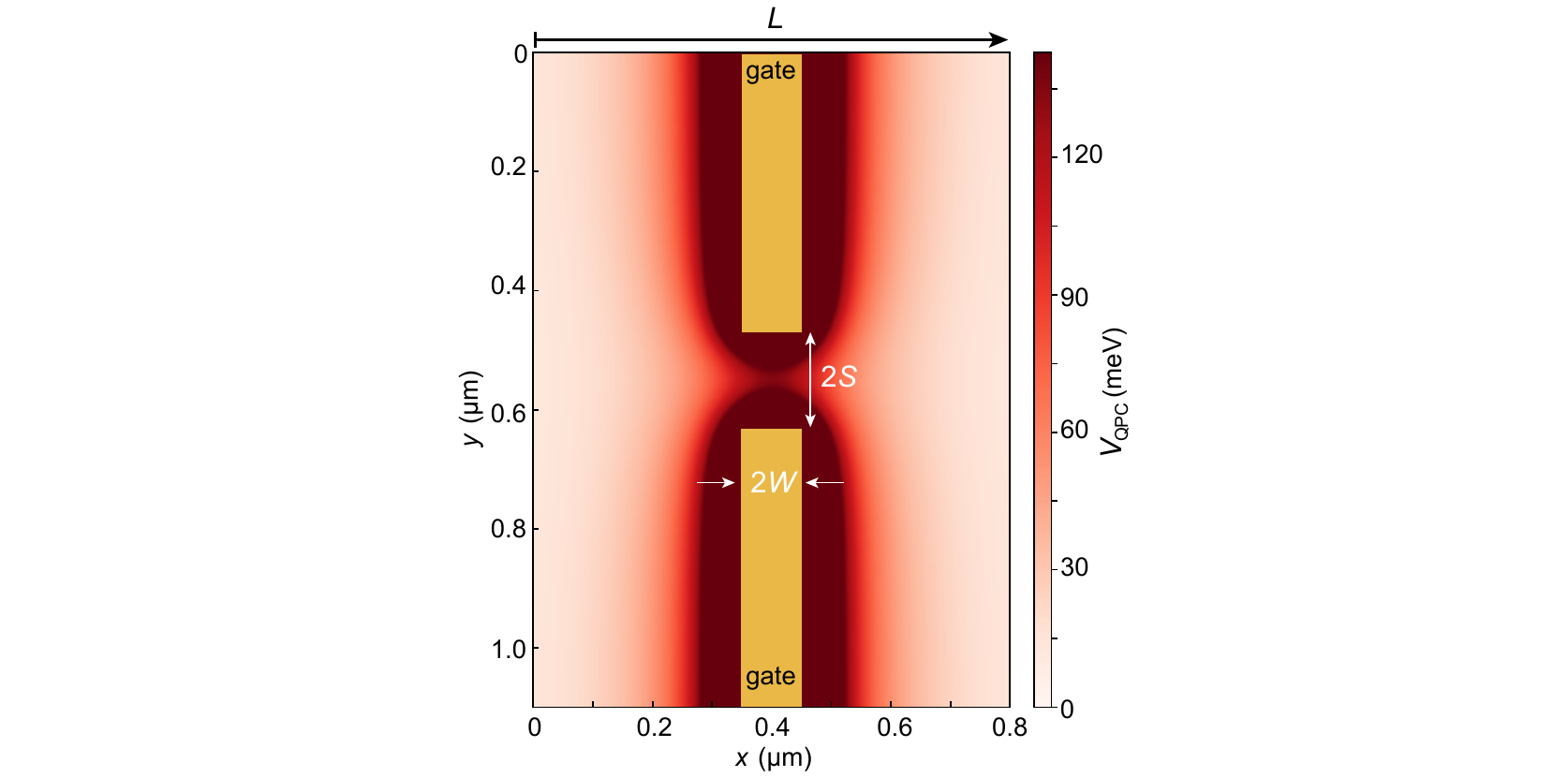}
\caption{\footnotesize{\textbf{QPC potential layout.}
The yellow contours show the geometry of the QPC gates and the red color depicts potential created at the position of 2DEG for $V_\text{g} = -1350$ mV.}}
\label{QPCpotential}
\end{figure}

For the simulation we adopt the following parameters: chemical potential $\mu = 143 $ meV, mean free path $l_\text{e} = 230$ nm, effective mass $m^*=0.05m_\text{e}$ (obtained from $k.p$ calculation of the Fermi velocity for a single mode quantum well in the growth direction). We also assume $T_\text{c} = 1.6$ K and $\Delta^*=190\;\mu$eV. The QPC geometry is set by the parameters: $W = 50$ nm (width of gates), $S=75$ nm (separation between gates) and $d = 50$ nm (distance from gates down to the 2DEG). 

We consider a system of the geometry similar to the one presented on Fig.~1b of the main text. Here the superconductor interface is located 230 nm after the QPC. 

\begin{figure}[h!]
\center
\includegraphics[width = 13cm]{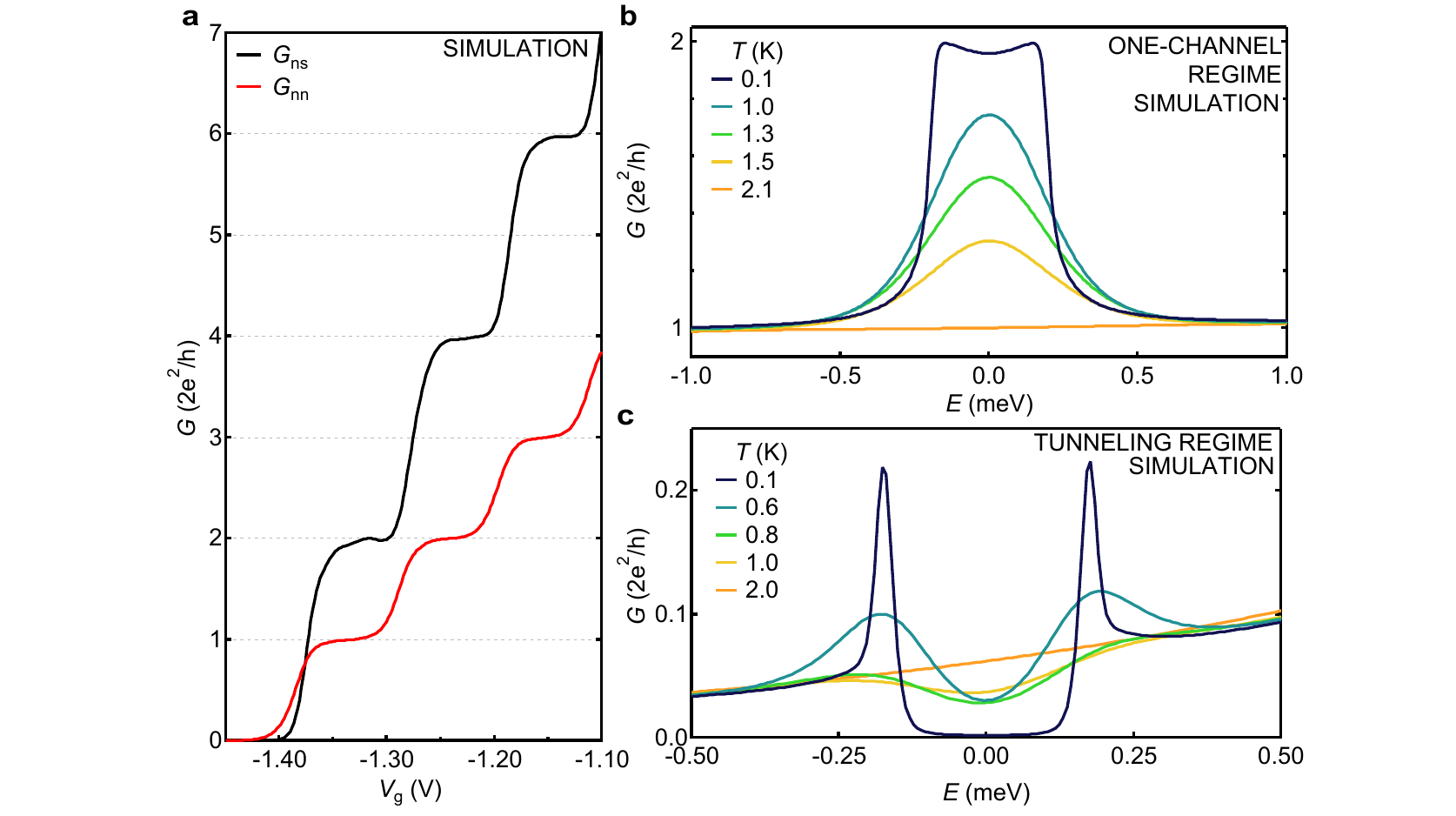}
\caption{\footnotesize{\textbf{Conductance calculated for a system with $L = 680$ nm.}
\textbf{a} Andreev-enhanced conductance $G_{ns}$ (black curve) and the normal-state conductance $G_{nn}$ (red curve) versus the potential on the QPC gates calculated for $E=0$.
\textbf{b} Spectroscopy curves in one-channel regime for $V_\text{g} = -1330$ mV.
\textbf{c} Tunneling spectroscopy curves for $V_\text{g} = -1408.7$ mV.
 }}
\label{short_system}
\end{figure}

Supplementary Figure \ref{short_system}a shows the conductance as a function of the gate voltage. The $G_\text{ns}$ conductance depicted with the black curve is quantized in multiplies of $4e^2/h$ as the transport involves transmission of an electron and an Andreev-reflected hole. Supplementary Figures \ref{short_system}b,c show the Andreev-enhanced spectroscopy curves obtained by varying the injection energy $E$. Supplementary Figures ~\ref{short_system}b and ~\ref{short_system}b show, respectively, the calculated finite--bias properties of the one--channel regime and the tunneling regime, for several values of the temperature. The value of $V_\text{g}$ in the simulations are chosen so the conductance at zero bias match the data at $T>T_\text{c}$ in Fig.~4 of the main text. The low temperature spectroscopy curves are similar the ones obtained by using the analytic expression of Blonder-Tinkham-Klapwijk (BTK) \cite{blonder_transition_1982}. However, for energies larger than the gap, the spectroscopy simulations show an increasing trend as a function of $E$ (cf. the orange curves on Supplementary Fig.~\ref{short_system}b,c where $T > T_\text{c}$), due to an increase of the energy of the injected particle with respect to the QPC potential. This dependence is pronounced in our geometry, because the slopes of the QPC steps are less than $50~$meV wide, making the conductance sensitive to changes in $E$ on the scale of single meV.

The low temperature one-channel spectroscopy curve shows maxima at $|E| \simeq \Delta$ (blue curve in Supplementary Fig.~\ref{short_system}b) while in the experimental data (cf. Fig.~4a and Fig.~5a of the main text) the curves decrease smoothly as $|V_\text{sd}|$ is increased. Previous theoretical work \cite{hekking_subgap_1994} showed that the detailed layout of the interface between the normal and superconducting electrodes (at the scale of the coherence length) impacts the subgap conductance due to interference between two electrons tunneling through the interface. Moreover, smearing of the superconducting coherence peak \cite{feigelman_universal_2012} is predicted to be an effect of disorder present in the superconducting film pointing again to the role of normal-superconductor interface.

The experimental structure consist of an extended 2DEG/superconductor interface created by the InGaAs/InAs heterostructure covered by Al. In the present calculations, we are limited to an abrupt semiconductor/superconductor interface. We therefore also consider a case where the distance from the QPC to the interface is increased relative to the lithographic dimensions.

\begin{figure}[h!]
\center
\includegraphics[width = 13cm]{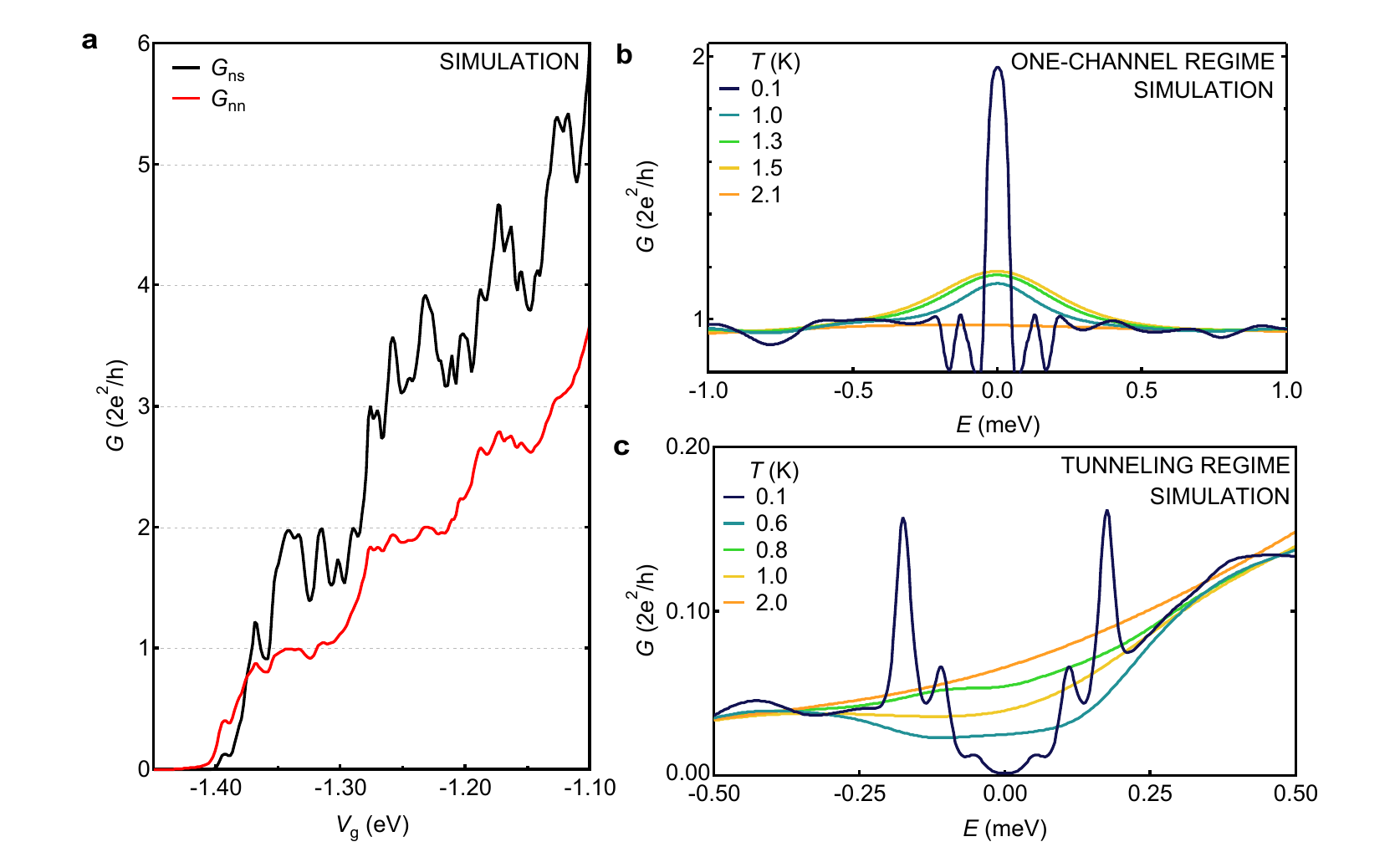}
\caption{\footnotesize{\textbf{Conductance calculated for a system with $L = 1250$ nm.}
\textbf{a} Andreev-enhanced conductance $G_\text{ns}$ (black curve) and the normal-state conductance $G_\text{nn}$ (red curve) versus the potential on the QPC gates calculated for $E=0$.
\textbf{b} Spectroscopy curves in the one-channel regime for $V_\text{g} = -1341$ mV.
\textbf{c} Tunneling spectroscopy curves for $V_\text{g} = -1407$ mV.
}}
\label{long_system}
\end{figure}
 
Supplementary Figure \ref{long_system} shows results obtained for a system with 800 nm distance between the QPC gates and the superconductor interface. In this calculation the scattering region is longer than the mean free path, leading to the peak/dip structures superimposed on the QPC conductance steps. Similar peaks/dips are observed in the experimental data in Figs.~2a,b of the main text. The fluctuations are more pronounced in the superconducting case ($G_\text{ns}$) due to the Andreev-enhanced conductance involving traversing the scattering region twice. The resonant features are also visible in the low temperature spectroscopy curves for energies larger the superconducting gap (cf. Supplementary Figs.~\ref{long_system}b,c), similar to the experimental curves in Figs.~4 a,c of the main text. Comparable pinch off curves are obtained when the disorder is located before the QPC, if the distance between the QPC and the superconductor are short.

The most notable feature of the system with extended length between the QPC and the superconductor is a significant reduction of the width of the central peak in the one--channel finite--bias simulations (blue curve in Supplementary Fig.~\ref{long_system}b). The rapid drop in conductance is a hallmark of an induced gap, for which the chaotic billiard in the region between the QPC gates and superconductor has zero density of states. The energy scale at which conductance drops is denoted $E_\text{b}$, and has the magnitude of Thouless energy \cite{melsen_induced_1996, melsen_superconductor-proximity_1997}, and hence it is inversely proportional to the area between the QPC and the interface. For $|E| > E_\text{b}$ the billiard has a non-zero discrete spectrum and so for $E_\text{b} < |E| < \Delta$ the conductance exhibits oscillations due to transport through resonant states which here are smoothed already for $T = 0.1$ K due to temperature averaging. The smooth resonances are also present in the low-temperature conductance curve in the tunneling regime (Supplementary Fig.~\ref{long_system}c).

\end{document}